# Allo-network drugs: harnessing allostery in cellular networks

Ruth Nussinov[1,2,1] , Chung-Jung Tsai[1], and Peter Csermely[3]

[1]*Center for Cancer Research Nanobiology Program, SAIC-Frederick, NCI-Frederick, Frederick, MD 21702, USA*
[2]*Sackler Institute of Molecular Medicine, Department of Human Genetics and Molecular Medicine, Sackler School of Medicine, Tel Aviv University, Tel Aviv 69978, Israel*, [3]*Department of Medical Chemistry, Semmelweis University, P O Box 260., H-1444 Budapest 8, Hungary*

**Abstract** Allosteric drugs are increasingly used because they produce fewer side effects. Allosteric signal propagation does not stop at the 'end' of a protein, but may be dynamically transmitted across the cell. Here, we propose that the concept of allosteric drugs can be broadened to *allo-network drugs*, whose effects can propagate either within a protein, or across several proteins, to enhance or inhibit specific interactions along a pathway. We posit that current allosteric drugs are a special case of allo-network drugs, and suggest that allo-network drugs can achieve specific, *limited* changes at the systems level, and in this way can achieve fewer side effects and lower toxicity. Finally, we propose steps and methods to identify allo-network drug targets and sites outlining a new paradigm in systems-based drug design.

**Running Title**: Allo-network drugs

**The chemical view and the network view of the cell are complementary**
Almost a decade ago, Hiroaki Kitano aptly noted "To understand biology at the system level, we must examine the structure and dynamics of cellular and organismal function, rather than the characteristics of isolated parts of a cell or organism" [1]. Kitano considered four systems properties: systems structures, dynamics, control and design. However, a mechanistic understanding of each property requires knowledge of the molecular structures, their associations, and their conformational dynamics. Together, these could provide an insight into why certain mutations can lead to disease, which on their own neither the knowledge of protein structures nor of protein-protein interaction networks may be able to achieve. Human perception often divides problems between scientific disciplines for convenience; however, such divisions do not exist in nature. Proteins, the key workhorse of the cell, are a case in point: the chemical view of protein molecules is of conformational ensembles; the network view perceives proteins as nodes in cellular networks. Chemists seek drugs against malfunctioning molecules; mathematicians and physicists consider the effects of drugs on networks of nodes and edges (interactions). Here we argue that the two views can be integrated, and this may expand the repertoire of drug targets.

**The concepts of allosteric drugs and cellular networks**
On the molecular level, orthosteric sites in related proteins are often similar because active sites evolved (or were preserved) to interact with similar ligands; consequently, drugs developed against one protein can often bind another. This is unlikely to be the case for allosteric drugs, because they bind on the protein surface away from the active site. Thus they can be more specific, and may have fewer side effects [2-7]. One example is the pleckstrin homology (PH) domain-dependent inhibitors of the phosphoinositide 3 kinase (PI3K)-Akt pathway, which exhibit selectivity and potency for individual serine/threonine kinase Akt isozymes [8]. Unlike the phosphatidylinositol (PI) analogs which inhibit Akt, but may also bind other PH domain containing proteins, the allosteric inhibitors of the individual Akt isozymes, such as MK-2206, appear selective for, and inhibit the activity of only Akt. In addition, because allosteric drugs do not block the active site and thus are not competitive with the natural substrate (ATP in this example), they do not act as an on/off switch for a receptor, and thus allow modulation of the levels of activity. Modulation can be beneficial. For example, small-molecule allosteric modulators of phosphodiesterase 4 (PDE4), the primary cAMP-hydrolyzing enzyme in cells, that do not completely inhibit enzymatic activity (Imax ~ 80-90%) have reduced potential to cause emesis, a dose-limiting side effect of existing active site PDE4 inhibitors, while maintaining biological activity. Such modulators of cAMP signaling can help central nervous system therapeutics where penetration to the brain is desired [9]. These advantages are reflected in the increasing number of allosteric drugs, from the age-old valium and the benzodiazepines to the more recent development of maraviroc (Selzentry) and cinacalcet (Sensipar), for AIDS and hyperparathyroidism, respectively [2-7]. Increasing numbers of pipelines are dedicated exclusively to the development of allosteric drugs. Recent examples of allosteric drugs include RDEA119/BAY 869766 as an allosteric mitogen-activated protein/extracellular signal-regulated kinase (MEK) inhibitor [10], pyrvinum, a potent allosteric inhibitor of Wnt signaling [11], efavirenz, an allosteric reverse transcriptase inhibitor [12] and SCF-I2, an allosteric inhibitor of (yeast) F-box protein Cdc4 [13].

---
[1] Correspondence: Ruth Nussinov, <u>ruthnu@helix.nih.gov</u>. Tel: 301-846-5579, fax: 301-846-5598.

To understand how an allosteric drug works, we need to consider proteins as ensembles of states, where each state has a certain population. A 'state' is a conformation, and the 'population' of the state is the concentration that it has in solution. The allosteric effect is the change in the populations (called 'population shift') of the states upon perturbations (e.g. ligand or drug binding, mutations) [14-17]. The conformation in which the active site is 'closed' is the most stable; consequently, it has the highest population in solution, and the protein shows high activity in the cell. The inactive conformation to which the allosteric drug binds has an 'open' active site. This conformation is less stable, and thus has lower population. However, upon binding of the drug to the allosteric site, the inactive conformation is stabilized, thus its population increases. This change in the relative distributions of the conformations in solution which takes place following binding is a 'population shift'. Figure 1 illustrates a schematic energy landscape showing a population shift between conformations upon allosteric drug binding, in this case for rapamycin/FKBP12 binding to the mTOR, and its consequences. Despite the clear advantages of allosteric drugs, allosteric designs do not always work: inhibitors can allosterically mimic a substrate, as in the case of a CD4 mimetic, which was designed to inhibit CD4 binding to gp120 to prevent gp120 interaction with CCR5 cell surface receptors and HIV-1 entry, but led to the opposite effect. Subsequent successful designs interrupted both CD4 and CCR5 interactions and achieved this goal [18]. Allosteric drugs also induced activation rather than repression in the phosphorylation of PKC and Akt [19-20]. Other problems can also surface [21]. In particular, the problem of the cell utilizing parallel pathways [22-23] is still a major hurdle [24]. These challenges emphasize the urgent need for additional novel strategies.

In the last decade, the analysis of the topology and dynamics of cellular networks (e.g., protein-protein interaction networks, signaling networks, gene transcription networks, metabolic networks) became a powerful tool to describe cellular functions in health and disease [17, 25-37]. Networks were increasingly used to aid the discovery of novel drug binding sites and drug targets [38-40]. However, to date, the two levels of description have been treated as largely distinct approaches.

**Broadening the concept of allosteric drugs to allo-network drugs**
Here, we propose that combining the chemical description of proteins as conformational ensembles with the system-level information related to cellular networks make it possible to broaden our view of allosteric drug discovery to the level of the network of the whole cell [41]. In the cell, all proteins are interconnected. Recent observations of network-wide propagation of allosteric effects in disease provide striking examples of network interconnectivity [42]: Alzheimer and prion diseases are neurodegenerative disorders, which are characterized by the abnormal processing of amyloid-β (Aβ) peptide and prion protein (PrPC), respectively. PrPc can either bind to Aβ oligomers or inhibit β-secretase 1 (BACE1) cleavage, and these events can lead to Aβ toxicity, long term memory potentiation and cell death. At the same time, the expression of PrPC itself appears to be controlled by the amyloid intracellular domain (AICD), which is obtained by cleavage of γ-secretase. PrPc transcription is regulated by p53 concentration, with p53 itself also under tight network control.

Here we argue that combining the chemical view with such network-level data may pave the way to novel generation of allosteric drugs which harness allostery at the cellular level. We name such drugs 'allo-network drugs'. Figure 2 illustrates a comparison of orthosteric, allosteric and allo-network drug action. Both the 'classical' allosteric drug (Figure 2B) and our proposed 'allo-network drug' (Figure 2C) are based on the principles of allostery and population shift (Figure 1). Further, both will have systems level effects and unbiased cell-based screening assays would pick up on such compounds. The difference is that the classical allosteric drugs are designed to directly target a specific (misfunctional) protein, as in the case of the GPCRs. By contrast, here we consider other proteins of the cellular network as final targets. Allo-network drugs integrate the chemical view of conformational ensembles with the network view of the cell. Such integration can be powerful because it allows a priori targeting of other proteins in the pathway, and in this way expands the scope of drug discovery targets. Allo-network drug targets may provide a large number of potential novel targets with high selectivity and fewer side effects. Thus the broadened new "allo-network drug" concept combines principles of allostery and systems level cellular network information. Within this broad description of allo-network drugs, current allosteric drugs are a special case.

Because allostery is an ensemble phenomenon [16], allosteric propagation can be described by a progressive change in the occupancy of states, that is, by the 'population shift' [16] (Figure 1). To understand how population shift takes place, consider the local perturbation that binding of a ligand creates: atoms which were in contact with the solvent or with neighboring atoms in the same molecule now interact with ligand atoms. To retain a favorable local atomic environment, a reorganization of the binding site takes place. This creates strain energy, and atoms around are also forced to change their contacts. In this way, the strain energy spreads and propagates in the structure through the dynamic fluctuations in atomic contacts leading to conformational and dynamic changes. Because allostery is the result of any structural perturbation, this will be the outcome of any binding, including orthosteric inhibition (Figure 2A). Allostery is a fundamental concept related to statistical thermodynamics [14-17, 43]. It takes place because proteins exist in ensembles of states around their native conformations, and the barriers that separate these are low (illustrated in Figure 1). Perturbation can change the relative populations of the states, and the time scales of the change relate to the barrier heights separating the energy minima of the states on the energy landscape. The perturbation energy (thus, conformational change) dissipates from the binding site through multiple pathways similar to waves generated by a stone thrown into the water. Further, binding to one protein leads to conformational and dynamic changes in adjacent proteins [17] across a pathway [44]. Packing in protein structures is not homogeneous; and consequently the propagation of the perturbation is not isosteric. Because cellular pathways consist of a series of events, allostery can explain how signals can propagate in the cell [44]. Thus, allosteric effects can be considered either at the level of the protein; or of the cell, where propagation can take place via large (in the case of some processes such as transcription initiation even gigantic) cellular assemblies over large (~hundreds or even thousands of Angstroms) distances (Figure 3).

**Examples of allo-network effects and drugs harnessing the principles of nature**
The fast emergence of cellular resistance to conventional and allosteric drug treatments necessitates expansion of the repertoire of drug targets and their combinations. Specific allosteric shape-shifting [4-5] at the functional site can be induced by binding far away. As an example, transcription factors (TFs) transfer signals from specific DNA regulatory elements (REs) via the gigantic Mediator complex [45] to initiate gene-specific transcription hundreds of Angstroms away (Figure 3A). For the glucocorticoid receptor (GR), a difference of one base pair in an RE sequence [46] or an amino acid mutation affects transcription initiation across such distances and leads to a functional change. Similarly, mutations in the p53 activation domain also led to such a distant specific effect [47]. By contrast, a mutation in Mediator, which often bridges between RE-bound TFs and the RNA polymerase II (Pol II) machinery and mediates transcription initiation, will affect transcription of many genes (Figures 3 A,B). In Figure 3A, a change in an RE, or mutational effects in the TF binding sites, may propagate via Mediator modules to Pol II and affect transcription initiation [46-47]. Perturbations caused by drug binding may also propagate similar to those of mutations or RE sequence changes, and may specifically interfere with signaling across such pathways. This may take place by either orthosteric or allosteric drugs (represented by $D_{11}$ and $D_{12}$, respectively, in Figure 3B). In both cases, the interaction between the protein and its neighbor along the pathway (TF1 and the Middle module of Mediator, in Figure 3B) may be affected, which may hinder the functional outcome (transcription activation in this case). However, these drugs do not affect transcription initiation by TF2 or TF3. By contrast, a drug that binds Pol II (Dp in Figure 3B) will inhibit transcription of all pathways going through Mediator.

This example illustrates how networks and allostery, which have been combined by evolution for cellular function [48], can be copied by allo-network drugs in design. Allo-network drugs can make use of the specifically wired cellular functional paths and the specific interactions of their perturbation channels. We argue that allo-network drugs whose effects inhibit (or enhance) specific interactions along a pathway would harness evolution; as such, they could achieve specific and *limited* systems changes, which could be advantageous for drug regimes.

**Protein disorder and modular organization help allo-network drug action**
Because allosteric effects involve release of strain energy through fluctuations and changes in van der Waals contacts and hydrogen bonds [49], propagation will be more efficient in a tightly packed environment with high atomic density. A mechanism ensuring tight packing at protein interfaces is the evolution of disordered proteins. Disordered proteins display a two-state behavior: stay unfolded in their unbound state, and become folded when another protein (or ligand) is bound. Their protein-protein interfaces can resemble the compact hydrophobic cores of single-chain proteins. By using disordered proteins, evolution engineered tight interfaces which can allow rapid long-range transfer of regulatory cues. The level of protein disorder has increased in human cells [50]. Disordered protein regions and the coupled large assemblies, along with the spatial functional modularity in the cell, amplify the allosteric relay [44]. The complexity of regulation has increased in higher organisms through binding of additional factors, as for example shown in the transcriptional Mediator assembly [45, 51]. There is a much larger number of signaling cross-talks in human cells than in *Drosophila* or *C. elegans* [52]. A recent study showed that denser networks are easier to regulate [53]. Thus, allo-network drug-like action occurs most where it is needed most: in human cells.

**Allo-network drugs and multi-drug targets as specific effectors of diseased cells**
Targeted therapies are considered to be less toxic and better tolerated than general chemotherapies because they target specific proteins and pathways. Promising allo-network drugs should interfere with disease-specific pathways. One example is the modulation of the PI3K pathway, a signaling network which is crucial to the growth and survival of many epithelial cancers: when a single targeted therapy was effective, the targeted proteins and pathways were observed to be master regulators of several pro-growth and pro-survival signaling pathways such as the PI3K/Akt, Raf/ERK, and Jak/Stat, which were all up-regulated in tumor cells [23]. An additional example relates to the HER2/EGFR and the c-Jun amino-terminal kinase (JNK)/c-Jun signaling pathways: the lapatinib inhibitor used in combination with anticancer agent capecitabine in targeted therapy against metastatic breast cancer, was found to also improve the proapoptotic effects of tumor necrosis factor-related apoptosis-inducing ligand (TRAIL) and two TRAIL receptor agonists in colorectal cancer [54]. Nonetheless, targeted therapies can also have side effects, such as cardiac toxicity in anti-HER2 therapy in combination with anthracyclines and inhibitors of angiogenesis. These again emphasize that the cellular network is inter-connected.

Combination therapies and multi-target drugs can provide a number of examples of allo-network drugs, because they combine multiple effects, often at places distant from the malfunctioning protein in the cellular network [25, 55]. As an example of allo-network multi-target action we describe the combination of MEK and B-RAF inhibition, which appears to be an efficient method to overcome tumor resistance against MEK inhibition therapy. Mutations typically arise to inhibitors targeting the active site. This can also take place against allosteric inhibitors, as was shown recently in COT, a serine/threonine kinase member of the MAPK family. Enhanced expression of COT activates the MAPK and the JNK pathways as well as stimulates the nuclear factor of activated T cells (NF-AT) and nuclear factor kappa B (NF-κB)-dependent transcription [56]. M307, a short-term culture derived from a B-RAF(V600E) tumor, developed resistance to the allosteric MEK inhibitor AZD6244 [57]. To uncover the possible mechanisms, clones resistant to the allosteric MEK inhibitor AZD6244 were also generated from two B-RAF(V600E) mutant colorectal cancer cell lines that were also highly sensitive to B-RAF inhibition [24]. The inhibition potency of the allosteric MEK inhibitor AZD6244 was restored by a B-RAF inhibitor. The combined efficiency of MEK and B-RAF inhibition, which overcame the resistance to MEK inhibitors, emphasizes the importance of knowledge at the cellular network level. The B-RAF inhibitor here may be regarded as a prototype of allo-network drugs acting at an adjacent protein, with MEK providing an efficient strategy to overcome the initial resistance. The serine/threonine kinase mammalian target of rapamycin (mTOR) provides a second example which also shows possible roles of compounds and signaling molecules (Figure 4). mTOR controls key cellular processes such as cell survival, growth and proliferation, and is frequently hyperactivated in a number of human malignancies [58-59]. mTOR is present in two distinct complexes, mTORC1 and mTORC2. Rapamycin/FKBP12 binds to mTOR and allosterically inhibits mTORC1 signaling but not to mTORC2, thus efficiently inhibiting some functions of mTOR but not all. However, mTORC2 is sensitive to prolonged rapamycin treatment which impedes mTORC2 assembly. Binding of rapamycin/FKBP12 to mTOR interferes with the ability of mSin1 and rictor to form mTORC2 and thus downstream signaling. Figure 4 illustrates how this example can provide insight into allo-network drugs. Increased activity of AKT and over-expression of S6K correlate with rapamycin activity.

**Possible steps and methods to identify allo-network drug targets**
Discovery of allo-network drug targets and sites may work as evolution; that is, via a trial-and-error strategy. Nonetheless, we argue that systems-level knowledge has reached a point which could allow higher efficiency of drug design. Below, we provide possible steps.
1. Mapping the network: The rapid increase in experimental structural and functional data combined with large scale modeling tools facilitate prediction of which proteins interact and how they interact [60]. The integration of the data and predictions helps the construction of the structural cellular pathways and of large, multi-molecular assemblies.
2. Analysis of the network to identify pathways: experimentally, this can involve for example, pathway profiling and cell-based pathway screening to characterize pathway selectivity for candidate inhibitors and specific signaling events, e.g., protein dislocation, degradation, secretion and expression; investigation of the interplay between signaling pathways; high-throughput RNAi screening to dissect cellular pathways and characterize gene functions; and using small molecules as probes of cellular pathways and networks, as for example in the Rho pathway in cytokinesis. Small molecules can be used not only as therapeutics to treat disease, but also as tools to probe complex biological processes [61]. Theoretical network tools can also help. For example, a bottom-up approach [38, 55] to simulate the superimposed action of allo-network drugs with disease-specific intracellular pathways; or top-down reverse engineering methods [62] to discriminate between 'high-' and 'low-intensity' pathways. In the case of directed networks, such as signaling or metabolic networks, we may construct tree-like hierarchies [63] to gauge the importance of various nodes, or find driver nodes controlling the network [53].
3. Identification of allosteric binding sites: experimentally, assays have been developed for detecting non-competitive modulators of protein-protein interactions targeting protein kinase A (PKA) [64]. A general computational strategy for identification of allosteric sites (e.g. $D_{12}$ in Figure 3B) may involve finding correlated motions between binding sites through examination of covariance matrix maps. Comparison of such maps can reveal which residue–residue correlated motions change upon ligand binding and thus can suggest new allosteric sites [65]. The allosteric SCF-I2 inhibitor of the SCF(Cdc4) ubiquitin ligase [13] provides experimental validation of an initial phase of this strategy.

Here, rather than tackling directly the flat protein-protein interfaces, we employ allostery through a pathway. Overall, detailed structural and functional knowledge and understanding of the pathways, the attributes and the information flow in the cellular network can offer guidelines to help circumvent the long trial-and-error process which was followed by evolution.

**Concluding remarks**
To allow cells to rapidly respond to changing conditions, evolution utilized compact interactions via conformational protein disorder, modular functional organization, and large multimolecular assemblies; all can facilitate dynamic signaling events over large distances. The allo-network drug concept we introduced here makes use of this functional attribute: it proposes to target malfunctioning proteins not only directly, but also via other proteins. Nature has shown that signaling can create a limited, specific, far away network functional change. We suggest that allosteric drugs can harness the same principles. Extending the repertoire of allosteric drugs by aiming at proteins other than the direct target could provide additional drug target candidates and help to reduce side effects. If those proteins share the target function, are in the same module, and are tightly bound (which allow efficient signal propagation), such a proposition could constitute an alternative drug strategy. We predict that methods utilizing structural and systems level knowledge to identify allo-network drug targets will significantly increase in the near future, and hope that these will lead to a next generation of medications with greater selectivity and effectiveness. Among these, key steps include (i) obtaining structures of proteins and their complexes and mapping them into low resolution EM density maps; (ii) relating the obtained structural maps to function to figure out key check points and allosteric propagation pathways; (iii) figuring out forward and backward loops; (iv) putting the structural pathways together to obtain a global picture of the cellular network in atomic resolution. These can reveal what goes wrong in disease, and possible toxic effects. Illustrating this by the mTOR pathway example, only if we have the structures of the mTORC1 and mTORC2 complexes and their interactions with subsequent components of the mTOR pathway, will we be able to understand its functional complexity, and why rapamycin/FKBP12 allosterically inhibits mTORC1 signaling but not mTORC2, while it is able to inhibit the assembly of the mTORC2 assembly over time (Figure 4) [58-59, 66]. Further, (v) Allosteric regulation of biological macromolecules is affected by both conformational and dynamic properties of the proteins and their complexes [67]. Trapping a protein in its inactive state [68] is expected to be a powerful strategy in allo-network drug discovery. This is because binding to a complementary conformation is faster. If the stability of that conformation is high, it will exist in solution in high population, which largely circumvents the need for a population shift and climbing over the barriers of the energy landscape (Figure 1). The challenge is to predict the conformational dynamics and the distributions of the states in solution, and even more so, the re-distribution following an allosteric event. To help this significant progress is being made using combination of spectroscopic techniques, particularly nuclear magnetic resonance (NMR), crystallography and molecular dynamics simulations [69]. Finally (vi), combining structural data with dynamics on the systems level under different conditions, such as mutational events [41], which can provide clues into alternative pathways in disease. This should also allow prediction of potentially harmful drug consequences. Despite the immense challenges, because such therapeutic development harnesses those same mechanisms which are used by nature, this should be possible.

**Acknowledgements** This project has been funded in whole or in part with Federal funds from the National Cancer Institute, National Institutes of Health, under contract number HHSN261200800001E, by the EU (FP6-518230; TÁMOP-4.2.2/B-10/1-2010-0013) and the Hungarian National Science Foundation (OTKA K69105 and K83314). The content of this publication does not necessarily reflect the views or policies of the Department of Health and Human Services, nor does mention of trade names, commercial products, or organizations imply endorsement by the U.S. Government. This research was supported (in part) by the Intramural Research Program of the NIH, National Cancer Institute, Center for Cancer Research.

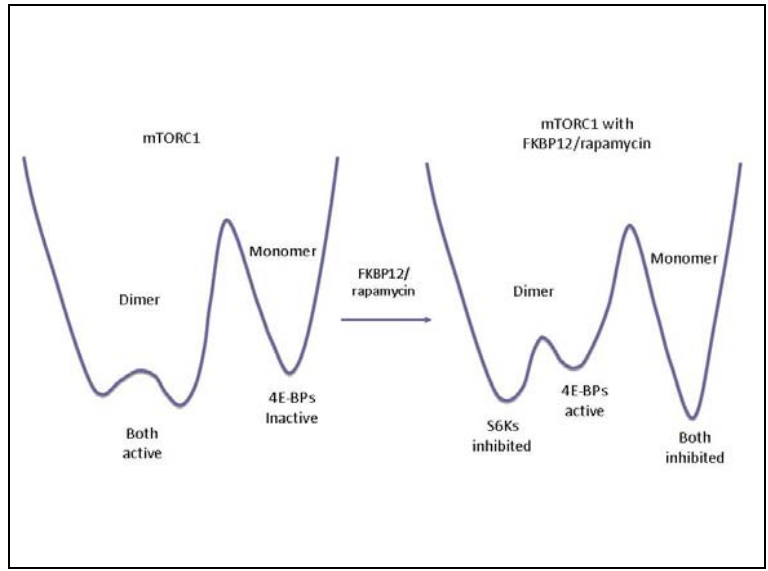

**Figure 1.** A schematic, simplified energy landscape which illustrates a population shift between conformations upon allosteric drug binding. The experimentally purified mTORC1 (mammalian target of rapamycin complex 1, for which there is a cryo-EM structure) system is used as an example. The figure reflects the changes in the activities of the two target proteins (S6Ks and 4E-BPs) after the FKBP12/rapamycin binds. In the drawing, the three free energy wells represent three conformational states of mTORC1. The conformation is labeled at the top of the well, and the activity is indicated at the bottom. Purified mTORC1 is shown by cryo-EM to have a preferred dimer conformation [66]. Therefore, a higher population time of the dimer (which is apparently the active form) in a lower free energy state is expected as depicted on the left. Upon FKBP12/rapamycin binding, S6Ks activity is inhibited without an appreciable delay; on the other hand, a time delay is observed for the 4E-BPs inhibition. Thus, in the landscape drawing (on the right), there is a low barrier between the two dimer conformations, "S6Ks inhibited" and "4E-BPs active", and a high barrier separates the dimer state from the monomer state. While in the purified system mTORC2 assembly is resistant to FKBP12/ramapycin and becomes sensitive only after a prolonged delay, in the cell the delay is not as long, probably because of involvement of additional factors. Figure 4 provides additional details.

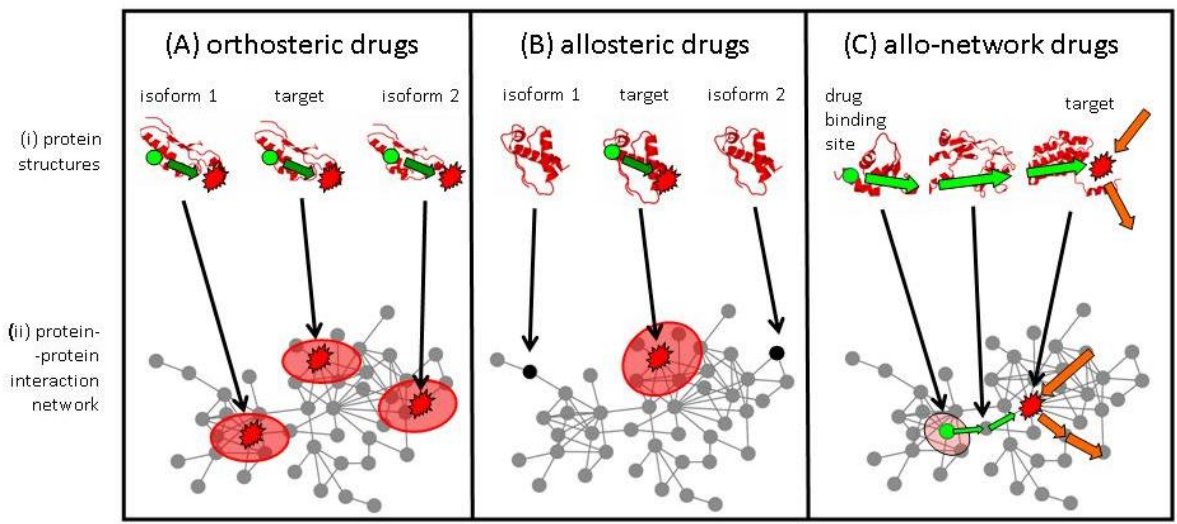

**Figure 2.** Comparison of orthosteric, allosteric and allo-network drugs. Here, in all cases the effect of the drug on the target site is via an allosteric propagation. Intra-protein propagation of drug-induced conformational changes is represented by dark green arrows. Conformational changes propagating through multiple proteins are marked with light green arrows. Drug binding sites are depicted by green circles; target sites are highlighted by red asterisks. **(A)** Orthosteric drugs. Here the inhibition (or activation, illustrated by light red ellipsoids at the bottom row) is via an allosteric effect which is elicited by active site (orthosteric) binding and propagates to a target site (dark green arrow). Because protein families often share similar binding pockets, orthosteric drugs can bind to multiple proteins (named here 'isoform 1' and 'isoform 2'), which can lead to side effects. **(B)** Allosteric drugs. Drug binding is in an allosteric site. Allosteric drugs are more specific than orthosteric drugs, because they usually do not bind to isoforms of the target. **(C)** Allo-network drugs. Here, drug binding is at an allosteric site; however, the target site is on a different protein in the cellular network. The pathway of allo-network drug-induced conformational changes (marked by light green arrows) may be highly specific and (or) specifically enhance (or inhibit) an intracellular pathway of propagating conformational changes (marked by orange arrows) at the target site. In promising allo-network drugs these intracellular pathways are disease-specific.

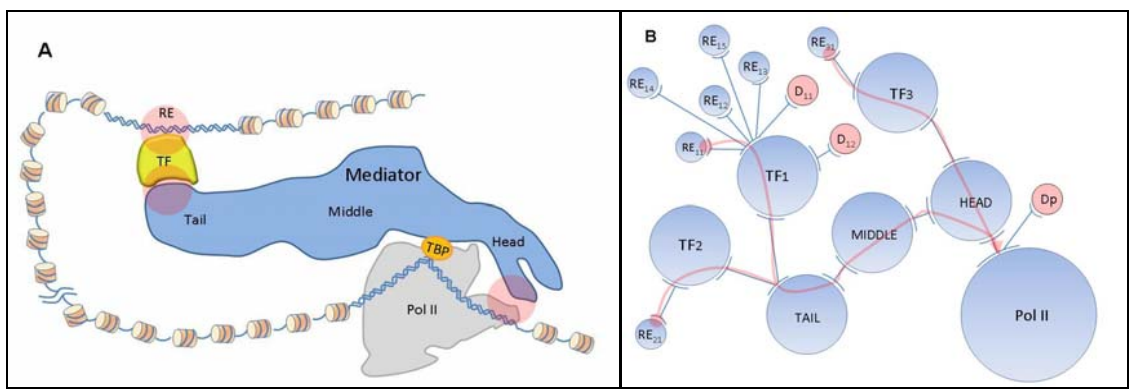

**Figure 3.** An example of specific, long-range allosteric propagation (on the order of hundreds or thousands of Angstroms) taken from nature. Here we propose that an allo-network drug can harness such regulatory pathways. The example involves gene-specific transcription initiation. **(A)** A schematic diagram illustrating transcription signal transduction from the DNA response element (RE), through transcription factor (TF) and the Mediator complex to the RNA polymerase II (Pol II) transcription machinery. For clarity, only the TATA-binding protein (TBP), Pol II and nucleosome-free DNA promoter are included in the drawing of the Pol II transcription machinery, instead of all the units involved in the pre-initiation complex (PIC). Between the RE and the Pol II transcription machinery, the DNA, which is covered with nucleosomes, is looped-out with variable lengths. While the TF is drawn with an arbitrary shape, the projected 2-D shape of Mediator and Pol II and their associations are based on cryo-EM structures. Mediator has three modules: Tail, Middle and Head. The long, multiunit-spanning transcription signal in the form of allosteric propagation starts from the RE via the TF, to the Tail, Middle, and Head modules of Mediator and finally down to Pol II. Three possible allo-network regulation sites are highlighted in pink circles. **(B)** A simplified transcriptional network. Only three transcription factors (TFs) and seven response elements (REs) are drawn here to illustrate transcriptional regulation in terms of allosteric propagations via REs, TFs and Mediator, down to RNA polymerase II (Pol II). Each TF has many REs and different TFs may bind different Mediator module (here Tail or Head), and has different propagation pathway. Each circle, a node of the network, represents either a segment of DNA (RE) or a group of proteins. Two arcs connected by a line represent either a protein-DNA or protein-protein interaction. An arc with many connected lines reflects the same interface of the node involved in interaction with different partners. The three pink arrows indicate three individual propagation pathways initiating at the interactions between REs and TFs and ending at the interface between Mediator Head and Pol II as depicted in Figure 3A. In terms of the *local network*, the drug $D_{11}$ that binds directly at the $TF_1$ DNA binding site is an *orthosteric drug* which prevents $TF_1$ from binding to $RE_{11}$, $RE_{12}$, $RE_{13}$, $RE_{14}$ and $RE_{15}$. On the other hand, the scenario where drug $D_{12}$ binds to $TF_1$ far from the DNA binding site where the resulting allosteric effect only prevents $TF_1$ binding to $RE_{12}$, labels $D_{12}$ as an *allosteric drug* whose function is to turn off the $TF_1$-$RE_{12}$ regulation. However, in terms of the *global network*, both drugs are allo-network drugs which alter all or some functions of $TF_1$ but do not perturb the function of $TF_2$ and $TF_3$ through the same Mediator-Pol II transcription regulation pathway, as compared to the drug Dp that binds Pol II at the interface with Mediator head thus shutting off all three TF regulatory functions.

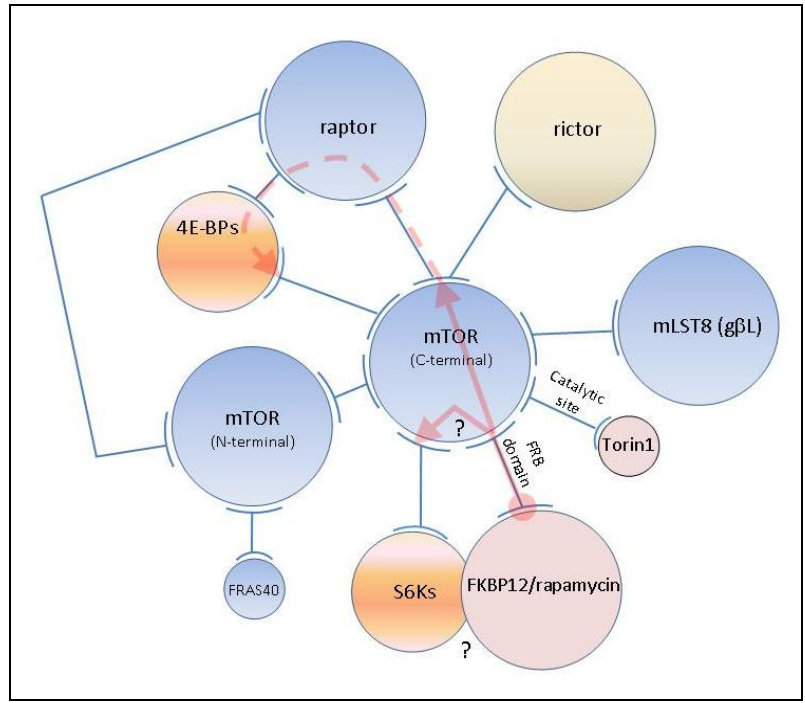

**Figure 4.** A schematic network of the mTORC1 and mTORC2 regulatory interactions. The network in the figure illustrates how FKBP12/rapamycin inhibits the phosphorylation of two target proteins, S6Ks (S6 kinase) and 4E-BPs (eukaryotic translation initiation factor 4E (eIF4E)-binding proteins), which are involved in the regulation of translation initiation. As in Figure 3B, each circle represents a domain, a protein or a group of proteins. Two arcs connected by a line represent protein-protein interaction. Circles in blue represent the assemblies of mTORC1. Rictor (a component of mTORC2) is in pale yellow, the drugs are in pink color, and the substrates of mTOR are colored in orange. The network representation is mainly based on the cryo-EM dimer structure of mTOR complex 1 (mTORC1) [66]. Upon binding of FKBP12/rapamycin to the FRB domain of mTOR, two possible allosteric propagation pathways (indicated by the pink arrows) can initiate at the interface between FKBP12/rapamycin and FRB: in the first, the propagation ends at the interaction of the mTOR C-terminal and the protein raptor in mTORC1 (or rictor, an exclusive partner of mTOR in the mTORC2 assembly; raptor and rictor appear to occupy the same surface on mTOR); in the second, the propagation ends at the interaction of mTOR C-terminal and the S6Ks substrate. It is not clear (as indicated by two question marks) whether the inhibition of S6Ks phosphorylation is because of the weakened recruitment of the second allosteric propagation or because of a steric exclusion as shown by the overlap between S6Ks and FKBP12/rapamycin complex. In mTORC1, the first allosteric propagation disrupts the association between raptor and mTOR C-terminal domain, which then prevents raptor from recruiting the substrate 4E-BPs in its (active) dimer-bound form, as indicated by the dashed pink arrow. Because raptor mediates the mTOR dimer interaction, now mTOR is a monomer. If we include the role of raptor in 4E-BPs phosphorylation, rapamycin is an allo-network drug. For clarity, the interactions between the substrates and the catalytic site of mTOR, at which the Torin1 drug binds, are not drawn.